\documentclass[runningheads]{llncs}

\usepackage[T1]{fontenc}
\usepackage{graphicx}
\usepackage{color}
\usepackage{physics}
\usepackage{wrapfig}

\spnewtheorem*{restatement}{Theorem}{\bfseries}{\itshape}

\begin{document}

\title{Robust Treasure Hunt in Anonymous Graphs with Quantum Pebbles by Oblivious Agents}
\titlerunning{Robust Treasure Hunt with Quantum Pebbles}

\author{
Gaurav Gaur\inst{1}
\and
Barun Gorain\inst{2}
\and
Rishi Ranjan Singh\inst{2}
\and
Daya Gaur\inst{3}
}
\authorrunning{G. Gaur et al.}

\institute{
 Department of Mechanical \& Industrial Engineering,
 University of Toronto, Toronto, Ontario M5S 3G8, Canada,
 \email{gauravgaur@mail.utoronto.ca}
 \and
 Department of Computer Science and Engineering,
 Indian Institute of Technology Bhilai, Bhilai, Chhattisgarh -- 491002, India,
 \email{\{barun,rishi\}@iitbhilai.ac.in}
 \and
 Department of Math and Computer Science,
 University of Lethbridge, Lethbridge, Alberta T1K3M4, Canada,
 \email{daya.gaur@uleth.ca}
}

\maketitle

% =====================================================================
\begin{abstract}
We study how to find a hidden treasure in anonymous graphs using an agent that
has no persistent memory. The nodes are indistinguishable, and  only edges have local port
numbers. Classical pebbles placed by an oracle cannot guide an oblivious agent to the treasure.
We introduce \emph{quantum pebbles}, which are sources that emit qubits in a fixed
(unknown) state, encoding at every node the outgoing port on the shortest
path to the treasure. By measuring
in several non-orthogonal bases, an oblivious agent recovers the port and
can reach the treasure in $D$ steps using $D$ quantum pebbles. This requires
$O(\Delta^{3}(\log D + \log \Delta))$ measurements per node, where $\Delta$ is
the maximum degree.

We further establish \emph{error robustness}, distinguishing two models of
state preparation error. Under \emph{per-node persistent} error, where a
device returns the same faulty encoding on every read, a single
mislabelled coloured pebble can trap an oblivious agent in an infinite loop and 
every randomized strategy decays exponentially in $D$. Quantum pebbles
inherit the same exponential decay. Under \emph{per-emission} error, the intended encoding is
correct, but each emitted qubit independently changes state as
$\rho = (1-e)\ketbra{\psi}{\psi} + e\sigma$ for an arbitrary noise matrix
$\sigma$. Here the quantum protocol is provably robust. A threshold decoding
rule with $O((\log D + \log \Delta)/\gamma^{2})$ measurements per basis, where
$\gamma = (1-e) - \delta_e$ and $\delta_e = (1-e)\delta + e$, has
success probability close to $1$ as $D \to \infty$, provided
$e < e^{*} = \sin^2(\pi/2\Delta)/(1+\sin^2(\pi/2\Delta))$. The separation that we establish is
thus among quantum pebbles with per-emission error and a persistent marker.

\end{abstract}

% =====================================================================
\section{Introduction}
\label{sec:intro}
% =====================================================================
Recent advances in the field of quantum computing invite a re-examination of classical problems in 
distributed computing. We study the well-known \emph{treasure hunt problem}. A
mobile agent starting at a source $s$ must locate a stationary treasure at
distance $D$ (known to the agent) in a connected, simple, undirected,
anonymous graph using a look--compute--move cycle. Nodes are unlabelled, and at each node $v$ the incident edges have
local port numbers $0,\dots,\deg(v)-1$ where $\Delta$ is the maximum degree.
The agent is oblivious, which ensures self-stabilization and
fault-tolerance \cite{flocchini2013computing}.
The full model is given in Section~\ref{sec:model}.

The treasure hunt in the deterministic setting has been studied using uniform
pebbles as a source of guidance placed by an oracle as permanent markers. There is an obvious
$O(D\Delta)$ algorithm for an agent with memory. The  pebbles are placed at each of
the $D-1$ internal nodes on the shortest path, and the agent remembers its
incoming port, scans the neighbours of each pebbled node to advance (at most $\Delta$ scans). For an
oblivious agent, this fails as the action of an agent at a node depends only on the presence or absence of a pebble, which does not determine which of
up to $\Delta$ ports leads to the treasure. Next, we state the two impossibility results,
whose proofs appear in Section~\ref{sec:impossibility}.

\newcommand{\stmtimposs}{For any pebble placement strategy $\Lambda$ using
 pebbles of one colour, it is not possible to design a deterministic treasure hunt
algorithm for use by an oblivious agent.}

\newcommand{\stmtcolours}{For any pebble placement strategy with at most
$\Delta-2$ coloured pebbles, it is not possible to design a deterministic treasure hunt
algorithm for an oblivious agent in a graph with maximum degree $\Delta$.}

\begin{theorem}[Uniform pebbles are insufficient]
\label{thm:impossibility}
\stmtimposs
\end{theorem}

\begin{theorem}[At least $\Delta-1$ colours are necessary]
\label{thm:colours}
\stmtcolours
\end{theorem}

A straightforward deterministic protocol does exist using $\Delta-1$ distinct
coloured pebbles (Section~\ref{sec:impossibility}). It is, however, \emph{fragile}: as
shown in Section~\ref{sec:robustness}, a mislabelled single-node can force an
oblivious agent into an  infinite loop. In contrast, the quantum
protocol developed here is robust under state preparation errors and measurement uncertainty through additional
sampling (Theorem~\ref{thm:quantum-robust}.

\subsection*{Our Contributions}
\begin{enumerate}
  \item We propose quantum pebbles as an information source, and
        an encoding of directional information into the fixed state of a single qubit
        (Section~\ref{sec:general}).
  \item We propose a strategy by which an oblivious agent finds the treasure in
        $D$ steps using $D$ quantum pebbles, with
        $O(\Delta^{3}(\log D + \log \Delta))$ measurements per node
        (Theorem~\ref{thm:main} and Remark~\ref{rem:delta-dependence}).
  \item \textbf{Robustness.} We give two error models. Under 
        persistent error at every node the classical $(\Delta-1)$-colour protocol fails after a
        single error and no randomized classical strategy beats exponential
        decay (Theorem~\ref{thm:exp-decay}). Quantum pebbles inherit the same
        bound (Proposition~\ref{prop:quantum-persistent}) for the per-node persistent error. Under per-emission
        error, the quantum protocol is robust up to an explicit threshold
        $e^{*}$ (Theorem~\ref{thm:quantum-robust}).
\end{enumerate}

\begin{table}[h]
\centering
\caption{Comparison of classical and quantum complexity for treasure hunt by
oblivious agents.}
\label{tab:complexity_comparison}
\begin{tabular}{|p{2cm}|p{4.5cm}|p{4.5cm}|}
\hline
\textbf{Metric} & \textbf{Classical} & \textbf{Quantum} \\
\hline
Feasibility & Impossible for uniform pebbles; feasible with $\Delta-1$ colours. & Feasible with quantum pebbles. \\
\hline
Search time & Random walks ineffective; little chance in $D^c$ steps. & $D$ steps (optimal). \\
\hline
%Encoding & $D$ pebbles of $\Delta-1$ colours. & Fixed quantum state of one qubit. \\
%\hline
Resource count & $\ge \Delta-1$ colours (deterministic). & $D$ quantum pebbles. \\
\hline
Cost & $1$-bit (or $\log\Delta$-bit) observation per round. & $O(\Delta^{3}(\log D+\log\Delta))$ measurements per node. \\
\hline
Robustness & Persistent error: fails on a single error. Per-emission error: not applicable. & Persistent error: same exponential decay. Per-emission error: succeeds w.p.\ $\to 1$ for $e<e^{*}$. \\
\hline
\end{tabular}
\end{table}

The remainder of the paper is organized as follows.
Section~\ref{sec:related} reviews the related work in brief. Section~\ref{sec:model} describes
the model. Section~\ref{sec:impossibility} is on the impossibility results for
classical pebbles. Section~\ref{sec:warmup} is the warm-up case
$\Delta=4$. Section~\ref{sec:general} presents the general quantum
construction and the main theorem. Section~\ref{sec:robustness} develops the
robustness results, and
Section~\ref{subsec:single-qubit} shows that encoding the entire path in a
single pebble is strictly worse than encoding one step per node. Finally, 
Sections~\ref{sec:discussion}--\ref{sec:conclusion} discuss extensions and
conclusions. A side-by-side comparison of classical and quantum approaches is 
summarized in Table~\ref{tab:complexity_comparison}.

% =====================================================================
\section{Related Work}
\label{sec:related}
% =====================================================================
In the area of mobile agents and mobile robots, the core challenges are search
and exploration. The influential book ``The Theory of Search Games and
Rendezvous'' by Alpern et al.~\cite{alpern2006theory} gives a comprehensive
overview of the challenges and seminal results on search problems.

The treasure hunt problem is a well-known variant of search and exploration in
anonymous graphs and has been extensively
studied~\cite{bouchard2023almost,gorain2022pebble,miller2015tradeoffs,pelc2021advice,BampasVolatilePheromone,komm2015treasure},
in both continuous~\cite{BhattacharyaPebble,langetepe2010optimality,bose2013revisiting}
and discrete~\cite{gorain2022pebble,xin2007faster} domains. A common formalism
is the notion of an \emph{algorithm with
advice}~\cite{gorain2022pebble,bouchard2020deterministic,pelc2021advice,pattanayak2024graph,BhattacharyaPebble},
which provides the agent with extra guidance such as angular hints, binary strings,
or pebbles to improve efficiency.

In the continuous domain, Bouchard et al.~\cite{bouchard2020deterministic}
established an optimal $O(D)$ bound with angular hints; Kao et
al.~\cite{kao1996searching} gave an optimal randomized algorithm on a line;
Pelc et al.~\cite{pelc2019cost,pelc2021advice} studied cost-versus-information
trade-offs in the plane and in geometric terrains;
Pelc~\cite{pelc2018reaching} treated the no-advice case.
Bhattacharya et al.~\cite{BhattacharyaPebble} examined pebbles-versus-time
trade-offs.

In the discrete domain, Miller et al.~\cite{miller2015tradeoffs} related prior
information to time for rendezvous and treasure hunt. Most relevant for the work here, Gorain
et al.~\cite{gorain2022pebble} studied treasure hunt in graphs with pebbles for
agents \emph{with} memory, giving an optimal $\Theta(D\log\Delta)$ algorithm.
Our work uses the same setting but for \emph{oblivious} agents using quantum
pebbles. 

Some recent papers \cite{akbaristoc25,Balliusoda26} have studied advantages of
quantum computing in the distributed setting. In \cite{Balliusoda26}, the
authors introduce quantum computing in a distributed network for locally
checkable labelling problems and show that the quantum-LOCAL model has improved
time complexity over the classical randomized-LOCAL model. In the other
direction, \cite{akbaristoc25} relates the quantum-LOCAL model to online
notions of locality, and thereby establishes limits on the quantum advantage
attainable for such problems.

Our approach is measurement-based. The idea of measurements in different bases is
also central to two universal models of quantum computation. The teleportation-based
model~\cite{nielsen2003quantum,nielsen1997programmable} stores and measures up
to four qubits at a time. Our protocol uses no unitaries before measurement and
no gate teleportation. The one-way
model~\cite{briegel2009measurement,raussendorf2001one} drives computation by
measuring a highly entangled cluster state. Our protocol uses no entanglement.
In our setting, measurements in different bases alone reveal the port to follow.

% =====================================================================
\section{Model and Preliminaries}
\label{sec:model}
% =====================================================================

\begin{definition}[Oblivious Agent]
\label{def:oblivious}
An agent is \emph{oblivious} if its internal state resets at the end of each
round. The computation depends solely on the observations made in the current round.
It cannot record visited nodes, traversed ports, or past observations.
\end{definition}

\begin{definition}[Deterministic Strategy]
\label{def:deterministic-strategy}
A \emph{deterministic strategy} $s$ is a function
\[
  s : \{1,\ldots,\Delta\} \times \mathcal{O}^{*} \longrightarrow \{0,\ldots,\Delta-1\},
\]
mapping the degree of the current node and the finite observation sequence
(pebble colours or measurement outcomes) to an outgoing port. Note that port
numbers are local. The same physical edge may carry different numbers at its
two endpoints, and the numbering has no global meaning. A strategy refers to port numbers, as the protocols of
Sections~\ref{sec:impossibility} and~\ref{subsec:gen-protocol} do when they
read a port number off a pebble colour or a measurement outcome.
\end{definition}

\begin{definition}[Randomized Strategy]
\label{def:randomized-strategy}
A \emph{randomized strategy} $\mathcal{S}$ is a probability distribution over
deterministic strategies; fresh coins are drawn at each node visit,
independently of history (the agent is oblivious).
\end{definition}

\begin{definition}[Quantum Pebble]
\label{def:qpebble}
A \emph{quantum pebble} is a source of qubits, all in the same, unknown, 
quantum state (the \emph{state of the pebble}). It is placed at a node by the
oracle and periodically emits a qubit in that state.
\end{definition}

\begin{remark}
\label{rem:not-two-colour}
A quantum pebble is not a two-coloured pebble: its state is unknown and only
revealed probabilistically by measurement, so the same pebble can yield
different outcomes when measured twice. 
%Emitting classical bits would instead
%require changing the emitted state, whereas every emitted qubit is identical.
\end{remark}
The agent repeats a \emph{look--compute--move} cycle in each round.
The objective is to locate the treasure in as few rounds as possible.

\begin{itemize}
\item \emph{Look.} At the current node $v$ the agent reads the degree
      $d = \deg(v)$ and, in the quantum setting, a finite sequence of classical
      outcomes obtained by measuring qubits emitted by the quantum pebble (Definition~\ref{def:qpebble}) at $v$
      across all $\lceil\Delta/2\rceil$ bases; beyond $D$ and $\Delta$, no global
      information is available.
\item \emph{Compute.} The agent evaluates a fixed strategy function on the
      observations of this round only
      (Definitions~\ref{def:deterministic-strategy}
      and~\ref{def:randomized-strategy}).
\item \emph{Move.} The agent traverses the edge with the chosen port number, or
      stays; at most one edge per round.
\end{itemize}

We adopt the following convention. If in some round the
computation does not determine a unique port (a \emph{decoding failure}),
the agent traverses port $0$, and the search is deemed unsuccessful, so every
such round is counted against the agent in all the bounds below.

The agent knows the distance $D$ to the treasure and the maximum degree
$\Delta$ of the graph; both are hard-wired into its strategy function and are
used to fix the number $n$ of measurements it performs per basis at a node
(Theorems~\ref{thm:special}, \ref{thm:main} and~\ref{thm:quantum-robust}). In
the error model of Section~\ref{sec:robustness}, we assume in addition that the
agent knows the preparation error rate $e$, or an upper bound
$\bar{e} \in [e, e^{*})$ on it. This is needed to set the decoding threshold of
Theorem~\ref{thm:quantum-robust}. This is the only global information available
to the agent. It still learns nothing
about the identity of the node it occupies, the distance it has travelled so
far, or the structure of the graph, and it retains no memory between rounds.
The impossibility results of Section~\ref{sec:impossibility} hold even for an
agent that knows $D$ and $\Delta$.

\subsection{Measurement Background}
\label{subsec:measurement-bg}
We use projective measurements in orthonormal bases; see
\cite{nielsen2010quantum} for background and notation. Measuring $\ket{\psi}=\alpha\ket{0}+\beta\ket{1}$ in the
computational basis $\{\ket{0},\ket{1}\}$ yields $0$ with probability
$|\alpha|^2$ and $1$ with probability $|\beta|^2$. In the Hadamard (sign) basis
$\{\ket{+},\ket{-}\}$, where $\ket{\pm}=\tfrac{1}{\sqrt2}(\ket{0}\pm\ket{1})$,
one observes $+$ with probability $\tfrac12|\alpha+\beta|^2$ and $-$ with
probability $\tfrac12|\alpha-\beta|^2$. Thus measuring the same state in two
bases distinguishes states: e.g.\ $\ket{\pm}$ measured in the bit basis gives
$0/1$ with equal probability but in the Hadamard basis gives a certain outcome.
The general $\lceil\Delta/2\rceil$-basis family that we need in the protocol is described in
Section~\ref{sec:gen-bases}.

\subsection{Computation Model}
\label{subsec:agent-model}
The agent is a \emph{memoryless, single-round automaton with a
quantum measurement oracle}. At each node, it observes the degree $d$ and a tuple
\[
  \mathbf{o} = \bigl(o_{j,\ell}\bigr)_{j \in \{0,\ldots,\lceil\Delta/2\rceil-1\},\; \ell \in \{1,\ldots,n\}}
  \in \mathcal{O}^{\,n\lceil\Delta/2\rceil},
\]
where $\mathcal{O}=\{j_{+},j_{-}: j=0,\ldots,\lceil\Delta/2\rceil-1\}$ are the measurement
outcomes and $n$
is the number of measurements per basis. Each $o_{j,\ell}$ comes from measuring
a fresh qubit in basis $M(j)$ (Section~\ref{sec:gen-bases}). The only
non-classical resource is projective measurement across
$\lceil\Delta/2\rceil$ bases; there are no quantum gates, no communication, and no
persistent storage. Given $(d,\mathbf{o})$ the agent evaluates $s$
(Definition~\ref{def:randomized-strategy}) and moves along the output port.

% ---------------------------------------------------------------------
\section{Impossibility Results for Classical Pebbles}
\label{sec:impossibility}
% ---------------------------------------------------------------------

\begin{figure}[ht]
\centering
\includegraphics[width=0.3\textwidth]{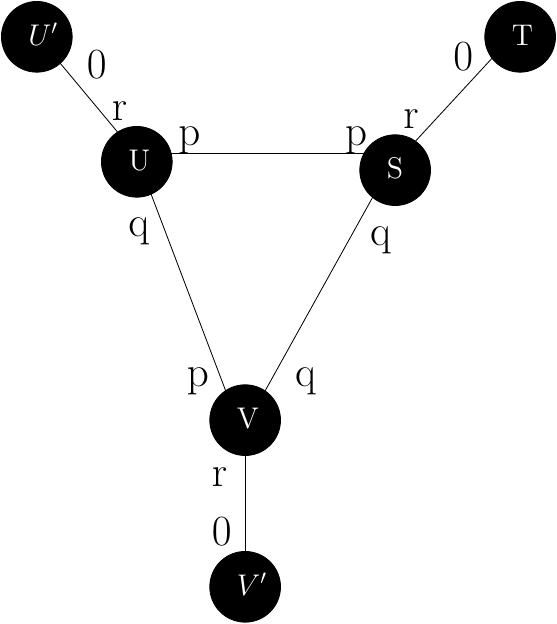}
\caption{The graph $G_{pqr}$, where $p,q,r \in \{0,1,2\}$ and distinct.}
\label{fig:NAVSC1}
\end{figure}

\begin{restatement}
\stmtimposs
\end{restatement}

\begin{proof}
Let \( p, q, r \in \{0, 1, 2\} \) be such that \( p \neq q \neq r \). We construct a graph \( G_{pqr} \) with a maximum degree \( \Delta = 3 \) as follows. Begin with a triangle over three nodes, \( S \), \( U \), and \( V \). To each of these nodes, we attach a degree-1 node. Connect \( T \) to \( S \), \( U' \) to \( U \), and \( V' \) to \( V \). We assign port numbers to each edge, as shown in Fig. \ref{fig:NAVSC1}. Let \( \mathcal{G} \) represent the class of graphs containing all the graphs \( G_{pqr} \) for different values of \( p \), \( q \), and \( r \).

Each graph \( G_{pqr} \) within \( \mathcal{G} \) is an instance of a treasure hunt where the agent starts at node \( S \) in \( G_{pqr} \), and the treasure is located at \( T \).

Since the agent is oblivious, its decisions at any given moment depend solely on the degree of the node it is at and whether a pebble is placed on that node. Hence the agent performs at most two distinct actions at a node of degree 3: one on seeing a pebble and one on seeing none. If either action is to stay at the current node, the agent never leaves \( S \) and so never reaches \( T \). Otherwise, let \( a \) and \( b \) be the ports taken on seeing a pebble and on seeing no pebble, respectively. Since \( \mathcal{G} \) contains a graph \( G_{pqr} \) for every choice of distinct \( p, q, r \in \{0,1,2\} \), it contains one in which the port numbers play the roles of \( a \) and \( b \) that the labels 1 and 0 play in \( G_{102} \). Without loss of generality, we may therefore assume that the agent will choose port 1 if it sees a pebble at a node with degree 3, and it will opt for port 0 if it does not see a pebble at a node with degree 3, and consider different cases in the graph \( G_{102} \) based on different placements of pebbles at the nodes.

\begin{enumerate}
\item A pebble is present at S: In this case, the agent moves to $U$. If a pebble is present at $U$, then in the next round, the agent again sees a pebble at a node with degree 3 and takes port 1 to reach $S$. This movement from $S$ to $U$ and then from $U$ to $S$ goes on indefinitely; hence, the agent can never find the treasure. If a pebble is not present at $U$, then it does not see a pebble at a node with degree 3 and hence takes port 0 and reaches node $V$. If a pebble is present at $V$, then the agent returns to $U$ via port 1 in the next round, and then the movement from $U$ to $V$ and then $V$ to $U$ goes on indefinitely; therefore, the agent can never find the treasure. Otherwise, the agent takes port 0 and returns to $S$, and the process is repeated. Hence, the agent can never reach the node $T$.
\item A pebble is not placed at $S$: In this case, using similar arguments as in Case 1, for every possible pebble placement at the node $U$ and the node $V$, the agent can never reach the treasure.
\end{enumerate}
Since the strategy was arbitrary, no deterministic algorithm for an oblivious agent succeeds on every instance in \( \mathcal{G} \).\qed \end{proof}

If pebbles with $\Delta-1$ different colours are available, the following
deterministic algorithm works. Let $s=v_0$, $T=v_D$, and
$P=v_0,v_1,\dots,v_D$ be the shortest path, with ports
$p_0,p_1,\dots,p_{D-1}$ where $v_i$ is reached from $v_{i-1}$ via $p_{i-1}$. Let
$\{c_1,\dots,c_{\Delta-1}\}$ be the colours. For each $i$ with $p_i\neq 0$, the
oracle places a pebble of colour $c_{p_i}$ at $v_i$. The agent moves along port
$0$ if it sees no pebble, and along port $i$ if it sees colour $c_i$. This
reaches the treasure in $D$ rounds. We now show $\Delta-1$ colours are
necessary.

\begin{restatement}
\stmtcolours
\end{restatement}

\begin{proof}
An instance of treasure-hunt is a tuple $<G,s,T>$, where $G$ is the underlying port-labelled graph, $s$ is the starting position of the agent, and $T$ is the position of the treasure. Assume $G$ is a star-shaped graph with the centre node $s$ and the leaf nodes $v_0\cdots v_{\Delta-1}$, where $s$ is connected to $v_i$ via port $i$. We create $\Delta$ different inputs $\mathcal{I}=\{<G,s,v_0>, <G,s,v_1>, \cdots, <G,s,v_{\Delta-1}>\}$.

Let $C=\{c_1, \cdots, c_{\Delta-2}\}$ be the set of colours of available pebbles. By $c_0$, we denote the scenario where no pebble is placed at a node. Since there are $\Delta-1$ (including $c_0$) possible ways a pebble can be placed at node $s$, while $\mathcal{I}$ contains $\Delta$ inputs, by the pigeonhole principle, there exist at least two distinct inputs in $\mathcal{I}$ for which the node $s$ receives the same pebble placement. Let these inputs be $<G,s,v_i>$ and $<G,s,v_j>$, where in both instances, the pebble of colour $c_k$ for $0\leq k\leq \Delta-2$ is placed at the node $s$. Upon seeing the same colour at the starting node, the agent must take the same action in both inputs. If that action is to stay at $s$, it never reaches the treasure in either input, so it must select the same port $p\in \{0, \cdots \Delta-1\}$ in both inputs $<G,s,v_i>$ and $<G,s,v_j>$. Since $i \ne j$, it follows that either $p\ne i$ or $p\ne j$. Without loss of generality, suppose that $p \ne i$. Then in $<G,s,v_i>$, the agent moves along the port $p$ from node $s$ to node $v_p$, where the treasure is not present. The only way to reach the treasure from $v_p$ is through node $s$. The agent must eventually return to $s$ in the subsequent step. Since the agent is oblivious, after reaching $s$, it again sees a pebble of colour $c_k$, moves along port $p$, and this process continues indefinitely. Hence, the agent can never reach the treasure in $<G,s,v_i>$.\qed \end{proof}

% ---------------------------------------------------------------------
\section{Warm-up: Graphs with $\Delta = 4$}
\label{sec:warmup}
% ---------------------------------------------------------------------
As a warm-up, we describe the quantum protocol for $\Delta=4$ and two bases, the
computational basis $\{\ket{0},\ket{1}\}$ and the
Hadamard basis $\{\ket{+},\ket{-}\}$. 
We use the \emph{local} port convention $1,2,3,4$ (rather than
$0,\dots,\Delta-1$). The oracle encodes port
$i\in\{1,2,3,4\}$ using the following states
\[
  \psi_1=\ket{0},\quad \psi_2=\ket{1},\quad \psi_3=\ket{+},\quad \psi_4=\ket{-}.
\]

\paragraph{Protocol.}
\begin{enumerate}
  \item The oracle places a quantum pebble in state $\psi_i$ at a node if the
        outgoing port on the shortest path is $i$. This uses $D$ quantum
        pebbles.
  \item At a node with pebble state $\psi_j$, the agent makes $n$ measurements
        in each of the two bases.
        \begin{enumerate}
          \item If $\psi_j\in\{\ket{0},\ket{1}\}$, the computational basis gives $0/1$
                with probability $1$ and the Hadamard basis gives $+/-$ with
                probability $1/2$ each and symmetrically for
                $\psi_j\in\{\ket{+},\ket{-}\}$.
          \item Exactly one basis is therefore expected to return $n$
                identical outcomes. The agent follows the port encoded by that
                outcome. Outcome $0 \to$ port $1$, $1 \to$ port $2$,
                $+ \to$ port $3$, and $- \to$ port $4$.
          \item If \emph{both} bases happen to return uniform strings of
                length $n$, the outcome is ambiguous and the search fails.
                Lemma~\ref{lemma:special} shows this happens with probability
                at most $1/2^{\,n-1}$ at a node.
        \end{enumerate}
  \item Step~2 repeats until the treasure is found or the failure condition is met.
\end{enumerate}

\begin{lemma}
\label{lem:special-encoding}
Let $O_1=0,\,O_2=1,\,O_3=+,\,O_4=-$. If the quantum pebble is in state
$\psi_i$, then exactly one of the computational-basis or Hadamard-basis measurements produces
outcome $O_i$ with probability $1$.
\end{lemma}
\begin{proof}
By Born's rule, measuring a state $\ket{\psi}$ in an orthonormal basis
$\{\ket{b_1},\ket{b_2}\}$ returns $b_m$ with probability
$\abs{\braket{b_m}{\psi}}^{2}$. For $\psi_1 = \ket{0}$ we have
$\abs{\braket{0}{0}}^{2} = 1$ and $\abs{\braket{1}{0}}^{2} = 0$, so the computational
basis returns $O_1 = 0$ with probability $1$; and since
$\ket{\pm} = \tfrac{1}{\sqrt2}(\ket{0} \pm \ket{1})$, we get
$\abs{\braket{\pm}{0}}^{2} = \tfrac12$, so the Hadamard basis returns each of
$+,-$ with probability $\tfrac12$. The computation for $\psi_2 = \ket{1}$ is
identical with $O_2 = 1$. For $\psi_3 = \ket{+}$ we have
$\abs{\braket{+}{+}}^{2} = 1$ and $\abs{\braket{-}{+}}^{2} = 0$, so the Hadamard 
basis returns $O_3 = +$ with probability $1$, while
$\abs{\braket{0}{+}}^{2} = \abs{\braket{1}{+}}^{2} = \tfrac12$, so the computational
basis returns each of $0, 1$ with probability $\tfrac12$; and similarly for
$\psi_4 = \ket{-}$ with $O_4 = -$. In each case exactly one of the two bases
returns $O_i$ with probability $1$, and the other returns either of its two
outcomes with probability $\tfrac12$. \qed
\end{proof}

\begin{lemma}
\label{lemma:special}
If the agent makes $O(\log D)$ measurements in each of the sign and bit bases
per node, the probability of finding the goal in $D$ steps is close to $1$.
\end{lemma}
\begin{proof}
The probability that state $\psi_i$ measured in the wrong basis $n$ times
produces a uniform sequence (all $n$ outcomes identical) is
$2\cdot(1/2)^{n} = 1/2^{n-1}$. 
Pebbles are prepared independently; bad events at distinct nodes are independent.
Choosing $n=O(\log D)$, the per-node success
probability is $1-1/D^{c}$ for a constant $c>1$. Using the same
$n$ at all $D$ nodes, the probability of reaching the goal is
$(1-1/D^{c})^{D} \ge 1 - D^{1-c}$, asymptotically close to $1$. \qed
\end{proof}

\begin{theorem}
\label{thm:special}
Let $\Delta = 4$. There is a strategy using $D$ quantum pebbles (one
per node) and $n = O(\log D)$ measurements per basis in each of the computational and
Hadamard bases, with which the agent finds the treasure in $D$ steps with probability at
least $(1-1/D^{c})^{D} \ge 1 - D^{1-c}$, where $c>1$ is the constant of
Lemma~\ref{lemma:special}. In particular, the success probability tends to $1$
as $D \to \infty$.
\end{theorem}
\begin{proof}
Lemma~\ref{lem:special-encoding} shows that the encoding is correct: exactly one
of the two bases yields the outcome $O_i$ with probability $1$, so a correct
decoding identifies the correct outgoing port. Lemma~\ref{lemma:special} bounds the
probability that the agent decodes every node on the shortest path correctly by
$(1-1/D^{c})^{D} \ge 1-D^{1-c}$ when $n = O(\log D)$ measurements are taken per
basis at each node. Combining the two, we have the claim. \qed
\end{proof}

% =====================================================================
\section{General Construction}
\label{sec:general}
% =====================================================================
The construction below generalizes the warm-up case $\Delta=4$ of
Section~\ref{sec:warmup}, where the bit and sign bases suffice and
Theorem~\ref{thm:special} locates the treasure in $D$ steps with $O(\log D)$
measurements per basis.
\subsection{Construction of Bases}
\label{sec:gen-bases}
Throughout this section we assume that $\Delta$ is even. This is no loss of
generality: the construction needs $\Delta$ distinct basis vectors to encode
$\Delta$ port numbers, and each basis supplies two, so an odd $\Delta$ is
handled by \emph{padding}. Use the construction below with $\Delta+1$ in place of
$\Delta$. This gives $\lceil \Delta/2 \rceil$ bases and $\phi = \pi/(\Delta+1)$, and
leave the one unused basis vector unassigned. This padding replaces $\delta$ by
$\cos^{2}(\pi/2(\Delta+1))$ in the subsequent proofs and changes none of the asymptotic
bounds. With this convention, the number of bases is $\lceil \Delta/2 \rceil$
in general. We write $\Delta/2$ below assuming even $\Delta$.
For $j \in \{0,1,\ldots,\Delta/2-1\}$ define
\[
  M(j) = \left\{ \tfrac{1}{\sqrt{2}}(\ket{0} + e^{ij\phi}\ket{1}),\;
                 \tfrac{1}{\sqrt{2}}(\ket{0} - e^{ij\phi}\ket{1}) \right\},
  \qquad \phi = \pi/\Delta .
\]
If $\phi=0$, this is the Hadamard basis for any $j$. The vectors of $M(j)$ are
orthonormal:
\[
  \tfrac{1}{2}(\bra{0} + e^{-ij\phi}\bra{1})(\ket{0} - e^{ij\phi}\ket{1})
  = \tfrac{1}{2}(\braket{0}{0} - e^{ij\phi}\braket{0}{1}
      + e^{-ij\phi}\braket{1}{0} - \braket{1}{1}) = 0 .
\]
Writing
\[
  \ket{j_+} = \tfrac{1}{\sqrt{2}}(\ket{0} + e^{ij\phi}\ket{1}),\qquad
  \ket{j_-} = \tfrac{1}{\sqrt{2}}(\ket{0} - e^{ij\phi}\ket{1}),
\]
any $\ket{\psi}=\alpha\ket{0}+\beta\ket{1}$ expands as
$\ket{\psi}=\braket{j_+}{\psi}\ket{j_+}+\braket{j_-}{\psi}\ket{j_-}$, and a
measurement in $M(j)$ returns $j_+$ (resp.\ $j_-$) with probability
$|\braket{j_+}{\psi}|^2$ (resp.\ $|\braket{j_-}{\psi}|^2$), collapsing the state. 
For $\Delta=4$ the angles for the two
bases $M(0),M(1)$ are shown in Table~\ref{tab:polar}.

%% Block sphere angles are off by a factor of two
%% this affects only the plot. rest is correct.
%Figure~\ref{fig:M0M1}) suffice.

%\begin{figure}
%\centering
%    \includegraphics[width=0.34\linewidth]{M0M1.png}
%    \caption{Vectors $\ket{0_+},\ket{0_-}$ (basis $M(0)$) and
%    $\ket{1_+},\ket{1_-}$ (basis $M(1)$) for $\Delta=4$, on the Bloch sphere.}
%    \label{fig:M0M1}
%\end{figure}

\begin{table}[h]
\centering
    \begin{tabular}{|c|c|c|}
        \hline
        $\ket{\psi}$ & $\theta$ & $\phi$ \\ \hline
         $\ket{0_+}$ &  $\pi/4$ & 0 \\
         $\ket{0_-}$ &  $2\pi - \pi/4$ & 0 \\
         $\ket{1_+}$ &  $\pi/4$ & $\pi/4$ \\
         $\ket{1_-}$ &  $2\pi - \pi/4$ & $\pi/4$ \\ \hline
    \end{tabular}
    \caption{Polar representation of basis vectors ($\Delta=4$),
    $\ket{\psi}=\cos\theta\ket{0}+\sin\theta\,e^{i\phi}\ket{1}$.}
    \label{tab:polar}
\end{table}

\subsection{Encoding and Protocol}
\label{subsec:gen-protocol}
Given the shortest path $v_1,\dots,v_{D+1}$ from $s$ to $t$ with outgoing ports
$e_1,\dots,e_D$, the oracle places at $v_i$ a quantum pebble with qubit in state $f(e_i)$,
where
\[
  f(j) = \begin{cases}
           \ket{i_+}, & j = 2i,\\
           \ket{i_-}, & j = 2i+1,
         \end{cases}
  \qquad i \in \{0,1,\ldots,\Delta/2-1\},\quad j \in \{0,1,\ldots,\Delta-1\}.
\]
Using this encoding, "one unknown qubit" can carry $\log_2\Delta$ bits of directional 
information without being prepared in $\Delta$ physically distinguishable states.
For example, on a path with ports $0,3,2,1,3,0$ and $\Delta=4$, the oracle
places pebbles emitting
$\ket{0_+},\ket{1_-},\ket{1_+},\ket{0_-},\ket{1_-},\ket{0_+}$.  At each node the
agent measures $n$ qubits in each basis $M(j)$. If exactly one basis yields $n$
identical outcomes $k^{*}$, the agent follows the port $f^{-1}(\ket{k^{*}})$;
by Lemma~\ref{lem:right-basis-prob1} the correct basis always yields identical
outcomes. If more than one basis yields a uniform run, decoding at this node
fails. The general
\emph{threshold rule} is stated in Section~\ref{subsec:robustness-thm}.

\begin{lemma}
\label{lem:right-basis-prob1}
If $\ket{j_+}$ (resp.\ $\ket{j_-}$) is measured in $M(j)$, the probability of
observing $j_+$ (resp.\ $j_-$) is $1$.
\end{lemma}
\begin{proof}
The probability of observing $j_{+}$ when the state is $\ket{j_{+}}$ is
\[
 |\braket{j_+}{j_+}|^2
 = \left|\tfrac12(\bra{0} + e^{-ij\phi}\bra{1})(\ket{0} + e^{ij\phi}\ket{1})\right|^2
 = 1 .
\]
Similarly, the probability of observing $j_-$ is $1$ when the state is
$\ket{j_-}$ measured in $M(j)$. The complementary event has probability $0$ by
orthonormality. \qed
\end{proof}

We write $*$ for either $+$ or $-$, so $\ket{i_*}$ is $\ket{i_+}$ or
$\ket{i_-}$. We use the following identities:
\begin{eqnarray}
    e^{i \theta} + e^{-i \theta} = 2\cos(\theta) \\
    \cos(2\theta) = 2\cos^2(\theta) -1 \\
    \cos(\theta) + \cos(-\theta) = 2\cos(\theta) \\
    \sin(\theta) + \sin(-\theta) = 0
\end{eqnarray}

\begin{lemma}
\label{lemma:angle}
For $j \neq k$ and $* \in \{+,-\}$,
$\ |\braket{j_*}{k_*}|^2 \le \cos^2\!\left(\dfrac{\pi}{2\Delta}\right).$
\end{lemma}
\begin{proof}
Here $M(j)$ with $j \neq k$ is called a \emph{wrong basis} for the state
$\ket{k_*}$, and $M(k)$ the \emph{correct basis}. There are four cases.

\emph{Case 1.}
\[
  \braket{j_+}{k_+}
  = \tfrac12(\bra{0} + e^{-ij\phi}\bra{1})(\ket{0} + e^{ik\phi}\ket{1})
  = \tfrac12(1 + e^{i(k-j)\phi}).
\]
Setting $(k-j)\phi=\theta$,
\[
  |\braket{j_+}{k_+}|^2
  = \tfrac14(1+e^{-i\theta})(1+e^{i\theta})
  = \tfrac12(1+\cos\theta),
\]
using $\cos(x)+\cos(-x)=2\cos(x)$ and $\sin(x)+\sin(-x)=0$. Assume $k>j$; then
$k-j\le \Delta/2-1$ and $\phi=\pi/\Delta$. Since $\cos$ is decreasing on
$[0,\pi/2]$, the largest value occurs at the smallest angle, so
\[
  |\braket{j_+}{k_+}|^2 \le \tfrac12\!\left(1+\cos\tfrac{\pi}{\Delta}\right)
  = \cos^2\!\left(\tfrac{\pi}{2\Delta}\right),
\]
using $\cos(2x)=2\cos^2(x)-1$.

\emph{Case 2.}
\[
  \braket{j_+}{k_-}
  = \tfrac12(\bra{0} + e^{-ij\phi}\bra{1})(\ket{0} - e^{ik\phi}\ket{1})
  = \tfrac12(1 - e^{i(k-j)\phi}),
\]
so $|\braket{j_+}{k_-}|^2 = \tfrac12(1-\cos((k-j)\phi))$. The largest value
occurs when $(k-j)\phi$ is largest, $\tfrac{\pi}{2}-\tfrac{\pi}{\Delta}$; using
$-\cos(x)=\cos(\pi-x)$,
\[
  |\braket{j_+}{k_-}|^2
  \le \tfrac12\!\left(1+\cos\!\left(\tfrac{\pi}{2}+\tfrac{\pi}{\Delta}\right)\right)
  \le \cos^2\!\left(\tfrac{\pi}{4}+\tfrac{\pi}{2\Delta}\right)
  \le \cos^2\!\left(\tfrac{\pi}{2\Delta}\right).
\]

\emph{Case 3.}
$\braket{j_-}{k_-}=\tfrac12(1+e^{i(k-j)\phi})$, identical to Case 1.

\emph{Case 4.}
$\braket{j_-}{k_+}=\tfrac12(1-e^{i(k-j)\phi})$, identical to Case 2.

In all four cases, the probability is at most
$\cos^2\!\left(\tfrac{\pi}{2\Delta}\right)$. \qed
\end{proof}

\begin{remark}
The bound of Lemma~\ref{lemma:angle} does not use the $\{\ket0,\ket1\}$ basis
employed in the $\Delta=4$ warmup (Section~\ref{sec:warmup}). This yields a
constant-factor difference in the  bound.
\end{remark}

\begin{remark}[Dependence on $\Delta$]
\label{rem:delta-dependence}
Since $1-\delta = \sin^{2}(\pi/2\Delta)$ and $-\ln(1-x)=x+O(x^{2})$, we have
$\log(1/\delta)=\Theta(1/\Delta^{2})$ and hence
$1/\log(1/\delta)=\Theta(\Delta^{2})$. Therefore, 
Theorem~\ref{thm:main} uses $O(\Delta^{2}(\log D+\log\Delta))$ measurements
per basis and $O(\Delta^{3}(\log D+\log\Delta))$ measurements per node. The
dependence on $\Delta$ has two sources: the $\Delta/2$ bases the agent must
try, and the fact that adjacent bases are only $\pi/\Delta$ apart, so
distinguishing them requires $\Omega(\Delta^{2})$ samples.

The corresponding cost under state preparation error is higher. In
Theorem~\ref{thm:quantum-robust} the relevant term is
$\gamma = (1-e)(1-\delta) - e$, which even in the noiseless limit $e \to 0$ is
only $1-\delta = \sin^{2}(\pi/2\Delta) = \Theta(1/\Delta^{2})$; hence
$1/\gamma^{2} = \Omega(\Delta^{4})$. So the threshold rule uses
$\Omega(\Delta^{4}(\log D + \log \Delta))$ measurements per basis and
$\Omega(\Delta^{5}(\log D + \log \Delta))$ per node, and $\gamma \to 0$ as
$e \to e^{*}$. The extra factor $\Delta^{2}$ relative to the noiseless case is
the cost of replacing an exact uniform run by a statistical test. Reducing it
is an open question.
\end{remark}

\begin{theorem}
\label{thm:main}
Let $\delta = \cos^2(\pi/2\Delta)$, which upper-bounds the probability of
observing any fixed outcome when measuring in a wrong basis. There is a
strategy using one quantum pebble per node ($D$ pebbles in total)
and a constant $c>0$ such that, for every number
\[
  n \;\ge\; c\,\frac{\log D + \log \Delta}{\log (1/\delta)}
\]
of measurements \emph{per basis}, the agent finds the treasure in $D$ steps
with probability at least $(1-\Delta\delta^{\,n})^{D} \ge 1-\Delta D\delta^{\,n}$.
Increasing the constant $c$ makes $\Delta D \delta^{\,n}$ arbitrarily small, so
the success probability can be brought arbitrarily close to $1$; the constant
$c$ controls how close. As the agent measures in each of the $\Delta/2$
bases, the number of measurements \emph{per node} is
$\tfrac{\Delta}{2}n = O\!\left(\Delta(\log D + \log \Delta)/\log(1/\delta)\right)$.
\end{theorem}
\begin{proof}
For state $\ket{k_*}$, call $M(k)$ the right basis and $M(j)$, $j\neq k$, a
wrong basis. In the right basis, every measurement returns $k_+$ or $k_-$. A
\emph{bad event} at a node is a length-$n$ uniform run ($j_+$ or $j_-$) in some
wrong basis, which leaves the port ambiguous between the wrong-basis reading and
the correct one. Let $\delta=\cos^2(\pi/2\Delta)$. By Lemma~\ref{lemma:angle}, a
run of $n$ copies of $j_+$ has probability $\le\delta^{n}$, and likewise for
$j_-$, so a fixed wrong basis produces a bad event with probability
$\le 2\delta^{n}$. With $\Delta/2-1$ wrong bases, the union bound gives total
bad-event probability $\le \Delta\delta^{n}$. Hence the agent decodes correctly
at a node with probability $\ge 1-\Delta\delta^{n}$, and over all $D$ nodes the
success probability is $(1-\Delta\delta^{n})^{D}$. Requiring
$\Delta\delta^{n}=O(1/D)$, i.e.\ $\Delta D \sim c(1/\delta)^{n}$, gives
$n = O((\log D + \log\Delta)/(\log 1/\delta))$, and choosing the constant so
that $\Delta\delta^{n} = c/D$ for arbitrarily small $c>0$ makes the success
probability, of locating the treasure, $(1-c/D)^{D} \ge 1-c$ arbitrarily close to $1$. \qed
\end{proof}

\section{Robustness}
\label{sec:robustness}
% =====================================================================
We study \emph{state preparation errors} in this section. The pebbles at each node (classical or
quantum) may be prepared incorrectly, independently at every node. We distinguish between
two different error models, and this distinction, rather than
the classical/quantum distinction \emph{per se}, is what determines whether
errors can be overcome when using quantum pebbles.

\begin{definition}[Per-node persistent error]
\label{def:persistent-error}
A guidance device (classical or quantum pebble) at $v_i$ has \emph{per-node persistent error}
$e \in [0,1)$ if, independently of all other nodes, the device encodes the
correct port $e_i$ with probability $1-e$, and otherwise encodes a port chosen
uniformly among the $\Delta-1$ incorrect ones. The encoding, once chosen, is
\emph{fixed}: every read of the device at $v_i$ results in the same (possibly
incorrect) port for classical pebbles (state for quantum pebbles).
\end{definition}

\begin{definition}[Per-emission error]
\label{def:emission-error}
A quantum pebble at $v_i$ has \emph{per-emission error} $e \in [0,1)$ if the
intended encoding $\ket{\psi_i}=f(e_i)$ is correct, but each emitted qubit
deviates independently, so that the qubit supplied on each read is the mixed
state $\rho_i = (1-e)\ketbra{\psi_i}{\psi_i} + e\,\sigma_i$ for an arbitrary
noise density matrix $\sigma_i$. This is a common model for state preparation
error for qubits.
\end{definition}

A classical coloured pebble is a persistent marker. It stores one symbol and
returns it on every read, so the only error model available to it is the
per-node persistent one. A quantum pebble, being an emitter, admits both. The
results below show that per-node persistent errors defeat both
paradigms, whereas per-emission errors can be averaged away by taking more
measurements.

\subsection{Persistent Errors}
\label{subsec:classical-fragility}

We show that under Definition~\ref{def:persistent-error} coloured and quantum
pebbles behave identically. The classical case is treated first.

\begin{example}[Single-node failure]
\label{ex:single-node}
Consider a graph with nodes $\{v_1,v_2,v_3\}$ and the path
$v_1 \to v_2 \to v_3$ with correct ports $(p_1 = 1, p_2 = 1)$, plus an extra edge
at $v_2$ whose port $p_2 = 2$ returns to $v_1$. Without error, the agent follows
port $1$ at $v_1$ to reach $v_2$, then port $1$ at $v_2$ to reach the treasure
at $v_3$. If a single error mislabels $v_2$ with port $2$, the oblivious agent
follows it back to $v_1$. Lacking any memory, it again follows port $1$ at $v_1$ and
enters the cycle $v_1 \to v_2 \to v_1 \to \cdots$. A single 
error causes unrecoverable failure.
\end{example}

\begin{definition}[Oblivious Adversary]
\label{def:oblivious-adversary}
An adversary is \emph{oblivious} if it commits to an error rate $e \in (0,1)$
before the agent chooses a strategy. Each pebble is then prepared independently
as in Definition~\ref{def:persistent-error}. The graph is $\Delta$-regular along $P$, and the correct outgoing
port $e_i$ is drawn uniformly from $\{0,\ldots,\Delta-1\}$ independently at each
node $v_i \in P$ by the adversary.
\end{definition}

\begin{lemma}[Single-step bound]
\label{lem:step-prob}
Fix a node $v_i$ of degree $\Delta \ge 2$ and let $e \in (0,1)$. Suppose that
\begin{enumerate}
  \item[(i)] the correct port $k = e_i$ at $v_i$ is uniform on
        $\{0,\ldots,\Delta-1\}$;
  \item[(ii)] the device at $v_i$ carries an \emph{encoded port} $m$, with
        $m = k$ with probability $1-e$ and $m$ uniform among the $\Delta-1$
        ports different from $k$ otherwise; and
  \item[(iii)] the device is the only source of information for the agent about the
        outgoing port at $v_i$. So that its action depends on the configuration at $v_i$
        only through the encoded port $m$. Formally, there is a kernel
        $q(\cdot \mid \cdot)$ with
        \[
          \Pr[\,\text{the agent traverses } p \mid k, m\,] \;=\; q(p \mid m)
          \qquad\text{for all } k, m, p .
        \]
\end{enumerate}
Then the probability of choosing the correct port at $v_i$ satisfies
\[
  p_{\mathrm{step}} \le \beta := \max\!\left(1-e,\;\tfrac{e}{\Delta-1}\right) < 1 .
\]
\end{lemma}
Hypotheses~(i) and~(ii) describe the adversary, while~(iii) notes a feature of
the model rather than a constraint on the agent. It says that the agent can learn about the correct port
$k$ only through the encoding $m$, and so rules out an agent
that could react to $k$ itself. Condition~iii) is automatic for both devices considered
below, since in our model the pebble is the only source of input at a node. 

\begin{proof}
Since the agent traverses at most one port in a round, 
$q(p \mid m)$ satisfies
\begin{equation}\label{eq:qnorm}
  \sum_{p=0}^{\Delta-1} q(p \mid m) \;\le\; 1
  \qquad\text{for every } m ,
\end{equation}
with equality precisely when the agent always moves. The inequality suffices
for arguments below. Any mass the agent places on staying at $v_i$ only lowers
the per-step success probability $p_{\mathrm{step}}$.
The probability in the definition of $q$ is taken over the random coins that the agent uses 
as well. So $q$ describes the behaviour after averaging over the
randomness. No structure needs to be assumed on $q$. A deterministic rule is the
special case where each $q(\cdot\mid m)$ puts all its mass on one port. Let 
$a_m := q(m \mid m)$, the probability that the agent follows the encoded port.

Let $A$ be the event that the agent traverses the correct port at $v_i$, so that
$p_{\mathrm{step}} = \Pr[A]$. Two random quantities of interest at the node are 
: the correct port $k$ and the encoded port $m$ carried by
the device. The
$\Delta^{2}$ pairs $(k,m)$ are mutually exclusive and exhaustive, so the law of
total probability, conditioning first on $k$ and then on $m$, gives
\[
  p_{\mathrm{step}}
  \;=\; \sum_{k=0}^{\Delta-1}\ \sum_{m=0}^{\Delta-1}
        \Pr[k]\;\Pr[m \mid k]\;\Pr[A \mid k,m] .
\]
Each factor corresponds to one hypothesis. By~(i) we have $\Pr[k] = 1/\Delta$,
and by~(ii) $\Pr[m \mid k] = 1-e$ when $m = k$ and $e/(\Delta-1)$ otherwise.
Finally
$A$ occurs precisely when the agent traverses $k$, so by~(iii)
$\Pr[A \mid k,m] = q(k \mid m)$. This is the one place where 
~(iii) is used,
replacing a quantity that could a priori depend on both $k$ and $m$ by one
depending on $m$ alone. Splitting the inner sum into the single term $m = k$ and the $\Delta-1$ terms
$m \neq k$ gives
\begin{equation}\label{eq:pstep1}
  p_{\mathrm{step}}
  \;=\; \frac{1}{\Delta}\sum_{k=0}^{\Delta-1}
        \Bigl[\,(1-e)\,q(k \mid k)
        \;+\; \frac{e}{\Delta-1}\sum_{m \neq k} q(k \mid m)\Bigr].
\end{equation}
%The first summand of~\eqref{eq:pstep1} is unchanged by renaming $k$ as $m$. In
%the second, the pairs $(k,m)$ with $k \neq m$ form a finite set of
%$\Delta(\Delta-1)$ elements, which the orders ``$k$ outer, $m \neq k$ inner''
%and ``$m$ outer, $k \neq m$ inner'' merely enumerate differently. 

Exchanging the two summations leaves the value unaltered, giving
\begin{equation}\label{eq:pstep2}
  p_{\mathrm{step}}
  \;=\; \frac{1}{\Delta}\sum_{m=0}^{\Delta-1}
        \Bigl[\,(1-e)\,q(m \mid m)
        \;+\; \frac{e}{\Delta-1}\sum_{k \neq m} q(k \mid m)\Bigr].
\end{equation}
By~\eqref{eq:qnorm}, $\sum_{k \neq m} q(k \mid m) \le 1 - a_m$, so
\begin{equation}\label{eq:pstep3}
  p_{\mathrm{step}}
  \;\le\; \frac{1}{\Delta}\sum_{m=0}^{\Delta-1}
        \Bigl[\,(1-e)\,a_m + \frac{e}{\Delta-1}\,(1-a_m)\Bigr]
  \;=\; \frac{e}{\Delta-1}
        + \bar{a}\Bigl[(1-e) - \frac{e}{\Delta-1}\Bigr],
\end{equation}
where $\bar{a} = \frac{1}{\Delta}\sum_m a_m \in [0,1]$. The right-hand side
of~\eqref{eq:pstep3} is affine in $\bar{a}$ and therefore attains its maximum
over $[0,1]$ at an endpoint: at $\bar{a}=1$ it equals $1-e$, and at
$\bar{a}=0$ it equals $e/(\Delta-1)$. Consequently
\[
  p_{\mathrm{step}} \;\le\; \max\!\left(1-e,\ \frac{e}{\Delta-1}\right)
  \;=\; \beta ,
\]
and $\beta < 1$ for every $e \in (0,1)$ and $\Delta \ge 2$. \qed
\end{proof}

\begin{remark}
\label{rem:follow-or-ignore}
When the agent always moves, \eqref{eq:pstep3} holds with equality. The two endpoints delimit what is achievable: $\bar{a}=1$ (always follow the pebble) achieves $1-e$, and
$\bar{a}=0$ (ignore it) achieves $e/(\Delta-1)$; every other strategy yields a
value between these two and so cannot exceed the better of them.
\end{remark}

In the coloured-pebble setting of Definition~\ref{def:persistent-error} the
agent observes the colour of the pebble, which \emph{is} the encoding; since the
pebble is the only input at $v_i$, its action there depends only on that
colour and hence only on $m$. Hypothesis~(iii) of
Lemma~\ref{lem:step-prob} is therefore automatic in our model, with
$q(p \mid m)$ the probability of traversing $p$ upon observing the colour that
encodes $m$, and hypotheses~(i) and~(ii) hold by
Definition~\ref{def:oblivious-adversary} and
Definition~\ref{def:persistent-error}. So Lemma~\ref{lem:step-prob} applies to
coloured pebbles.

\begin{theorem}[Exponential decay]
\label{thm:exp-decay}
For any randomized strategy and any $e\in(0,1)$, there are instances on which
\[
  \Pr[\text{agent finds treasure within } D \text{ steps}] \le \beta^{D},
  \qquad \beta=\max\!\left(1-e,\tfrac{e}{\Delta-1}\right)<1.
\]
\end{theorem}
\begin{proof}
Consider an instance as in Definition~\ref{def:oblivious-adversary}, that is,
$\Delta$-regular along $P$ and with a uniformly random labelling, and in which
the shortest path $P=v_1,\dots,v_D$ to the treasure is unique; then to succeed
within $D$ steps the agent must take the correct port at each of
$v_1,\dots,v_D$ in order. Let $\mathcal{E}_i$ be the event that the agent
reaches $v_i$ and steps correctly there. Then
\[
  \Pr\!\left[\bigcap_{i=1}^{D}\mathcal{E}_i\right]
  = \prod_{i=1}^{D}\Pr[\mathcal{E}_i\mid \mathcal{E}_1,\dots,\mathcal{E}_{i-1}].
\]
Conditioned on reaching $v_i$, the correct-step probability is at most $\beta$
by Lemma~\ref{lem:step-prob}, because (i) pebble errors are
independent across nodes (Definition~\ref{def:oblivious-adversary}) and (ii) the
 decision at $v_i$ depends only on the colour at $v_i$ and its
fresh coins. Hence
$\Pr[\text{success within } D \text{ steps}] \le \prod_{i=1}^{D}\beta = \beta^{D}$,
which decays exponentially since $\beta<1$. \qed
\end{proof}

\begin{remark}
Even against an oblivious adversary and with optimal random strategy, the
coloured-pebble approach has $D$-step success probability vanishing
exponentially.
\end{remark}

Quantum pebbles whose preparation is persistently wrong also fail in exactly the
same way.

\begin{proposition}[Quantum pebbles under persistent error]
\label{prop:quantum-persistent}
Suppose each quantum pebble is subject to per-node persistent error
$e \in (0,1)$ in the sense of Definition~\ref{def:persistent-error}. The pebble
at $v_i$ emits qubits in state $f(e_i')$, where $e_i' = e_i$ with
probability $1-e$ and is otherwise a port drawn uniformly among the $\Delta-1$
incorrect ones, and \emph{every} emission at $v_i$ is for the corresponding encoding. Then
on the instances of Theorem~\ref{thm:exp-decay}, for any strategy of the
oblivious agent and for \emph{every} number of measurements per basis,
\[
  \Pr[\,\text{agent finds treasure within } D \text{ steps}\,]
  \;\le\; \beta^{D},
\]
where $\beta = \max(1-e,\,e/(\Delta-1)) < 1$ is as in
Theorem~\ref{thm:exp-decay}.
This is exactly the bound of Theorem~\ref{thm:exp-decay}: under persistent
error, quantum pebbles offer no advantage whatsoever over coloured pebbles.
\end{proposition}

\begin{proof}
We apply Lemma~\ref{lem:step-prob}, taking the encoded port at $v_i$ to be
$e_i'$. Hypotheses~(i) and~(ii) hold by
Definition~\ref{def:oblivious-adversary} and
Definition~\ref{def:persistent-error}, as in the classical case, so only
hypothesis~(iii) has to be checked.

Every qubit the pebble at $v_i$ supplies is in the same state $f(e_i')$. Each
measurement outcome is therefore governed by $e_i'$ and the basis used, so the
outcomes the agent collects at $v_i$, and with them the port it finally
traverses, have probabilities that depend on $e_i'$ alone. Since the pebble is
the only input at $v_i$, nothing else there can inform it about the
correct port, so hypothesis~(iii) holds, with
\[
  q(p \mid m) \;=\; \Pr[\,\text{the agent traverses } p \mid e_i' = m\,] ,
\]
and Lemma~\ref{lem:step-prob} gives $p_{\mathrm{step}} \le \beta$.

The number $n$ of measurements plays no role. The argument of
Theorem~\ref{thm:exp-decay} (errors independent across nodes, agent oblivious)
then gives $\prod_{i=1}^{D}\beta = \beta^{D}$. \qed
\end{proof}

\begin{remark}[What persistence costs]
\label{rem:persistence}
Under a
persistent error every read reports the same corrupted state, so the error is
frozen and no averaging is possible. The advantage established in
Theorem~\ref{thm:quantum-robust} below is therefore not an advantage of
quantum over classical devices as such, but of a \emph{regenerative emitter}
over a \emph{persistent marker}.
\end{remark}

\subsection{Per-Emission Error}
\label{subsec:quantum-error}
In this section we use the density matrix notation,
see \cite{nielsen2010quantum}, and recall $\phi=\pi/\Delta$, the bases $M(j)$,
and the encoding $f$ from Section~\ref{sec:gen-bases}. Throughout the
subsection, the error model is the per-emission one of
Definition~\ref{def:emission-error}, and $\rho_i$ denotes the mixed state
defined there, which the pebble at $v_i$ supplies on each read.

We write $\ket{k^{*}}$ for $\ket{k_+}$ or $\ket{k_-}$. $M(k)$ is the correct basis, and any $M(j)$, $j\neq k$, is a wrong basis.

\begin{lemma}[Correct-basis statistics under error]
\label{lem:correct-basis}
For a qubit emitted with per-emission error $e$,
\[
  \Pr[\text{observe } k^{*}\mid \text{correct basis } M(k)]
  = \Tr(\ketbra{k^{*}}{k^{*}}\rho_i) \ge 1-e.
\]
\end{lemma}
\begin{proof}
By linearity of the trace,
\[
  \Tr(\ketbra{k^{*}}{k^{*}}\rho_i)
  = (1-e)\underbrace{|\braket{k^{*}}{\psi_i}|^{2}}_{=1}
  + e\,\Tr(\ketbra{k^{*}}{k^{*}}\sigma_i)
  \ge (1-e)\cdot 1 + e\cdot 0 = 1-e,
\]
using $|\braket{k^{*}}{\psi_i}|^{2}=1$ (Lemma~\ref{lem:right-basis-prob1}) and
non-negativity of the trace. \qed
\end{proof}

\begin{lemma}[Wrong-basis statistics under error]
\label{lem:wrong-basis}
Let $\delta=\cos^2(\pi/2\Delta)$. For $j\neq k$,
\[
  \Pr[\text{observe } j^{*}\mid \text{wrong basis } M(j)]
  \le \delta_e := (1-e)\delta + e.
\]
\end{lemma}
\begin{proof}
By linearity of the trace,
\[
  \Tr(\ketbra{j^{*}}{j^{*}}\rho_i)
  = (1-e)\underbrace{|\braket{j^{*}}{\psi_i}|^{2}}_{\le\delta}
  + e\,\underbrace{\Tr(\ketbra{j^{*}}{j^{*}}\sigma_i)}_{\le 1}
  \le (1-e)\delta + e = \delta_e,
\]
using $|\braket{j^{*}}{\psi_i}|^{2}\le\delta$ for $j\neq k$
(Lemma~\ref{lemma:angle}). \qed
\end{proof}

\begin{remark}[Distinguishability threshold]
\label{rem:distinguishability}
The agent can statistically distinguish $M(k)$ from a wrong $M(j)$ iff the
probability of the correct outcome in the correct basis exceeds that of the same
outcome in a wrong basis, i.e.\ $1-e > (1-e)\delta+e$. Rearranging,
$1-\delta > e(2-\delta)$, giving the tolerance threshold
\[
  e < e^{*} = \frac{1-\delta}{2-\delta}
            = \frac{\sin^{2}(\pi/2\Delta)}{1+\sin^{2}(\pi/2\Delta)}.
\]
%where $1-\delta=\sin^2(\pi/2\Delta)$ and $2-\delta=1+\sin^2(\pi/2\Delta)$.
\end{remark}

\subsection{Robustness Guarantee for Per-Emission Errors}
\label{subsec:robustness-thm}

The uniform-run rule of
Section~\ref{subsec:gen-protocol} is no longer adequate. By
Lemma~\ref{lem:correct-basis}, even the correct basis produces $n$ identical
outcomes with probability only about $(1-e)^{n}$, which vanishes as $n$ grows.
The agent therefore uses the following \emph{threshold rule}. Fix
\[
  \gamma := (1-e) - \delta_{e},
  \qquad
  \tau := \delta_{e} + \tfrac{\gamma}{2} = \tfrac{(1-e)+\delta_{e}}{2}.
\]
By Remark~\ref{rem:distinguishability}, $\gamma > 0$ precisely when
$e < e^{*}$. The threshold rule is the general form of the decoding step, and
the uniform-run rule of Section~\ref{subsec:gen-protocol} is its
specialization to $e = 0$. There $\delta_{0} = \delta$ and
$\gamma = 1-\delta > 0$, and by Lemma~\ref{lem:right-basis-prob1} the correct
basis attains empirical frequency exactly $1$. For each basis $M(j)$, the agent computes the empirical frequency of
each of the two outcomes over its $n$ measurements. It follows the port
encoded by an outcome whose empirical frequency is at least $\tau$; if no
outcome, or more than one outcome, reaches the threshold, decoding at this node
fails. Computing $\tau$ requires the agent to know $e$, which we assume as part
of the model. Knowledge of an upper bound suffices. If the agent knows only
some $\bar{e} \in [e, e^{*})$ and uses
$\bar{\gamma} := (1-\bar{e}) - \delta_{\bar{e}} > 0$ and
$\bar{\tau} := \delta_{\bar{e}} + \bar{\gamma}/2$ in place of $\gamma$ and
$\tau$, then, because $\delta_{e} = \delta + e(1-\delta)$ is increasing in $e$
while $1-e$ is decreasing, we still have
$1-e \ge 1-\bar{e} = \bar{\tau} + \bar{\gamma}/2$ and
$\delta_{e} \le \delta_{\bar{e}} = \bar{\tau} - \bar{\gamma}/2$. Both gaps
remain at least $\bar{\gamma}/2$, so Theorem~\ref{thm:quantum-robust} holds
verbatim with $\bar{\gamma}$ in place of $\gamma$.

\begin{theorem}[Robustness under per-emission error]
\label{thm:quantum-robust}
Assume each quantum pebble is subject to per-emission error $e$ in the sense of
Definition~\ref{def:emission-error}, with the intended encoding correct at
every node. Let $\delta=\cos^2(\pi/2\Delta)$ and
$e<e^{*}=\tfrac{\sin^{2}(\pi/2\Delta)}{1+\sin^{2}(\pi/2\Delta)}$, and set
$\delta_e=(1-e)\delta+e<1$ and $\gamma = (1-e)-\delta_{e} > 0$. There is a
constant $c>0$ such that, if at each node the agent applies the threshold rule
with any number
\[
  n \;\ge\; c\,\frac{\log D + \log\Delta}{\gamma^{2}}
\]
of measurements per basis, then
\[
  \Pr[\text{succ}] \ge \left(1-\Delta e^{-n\gamma^{2}/2}\right)^{D}
  \longrightarrow 1
  \quad\text{as } D\to\infty.
\]
The number of measurements per node is $\tfrac{\Delta}{2}n$.
\end{theorem}
\begin{proof}
Fix a node $v_i$ with intended encoding $\ket{k^*}$. The measurements are
independent, since each is performed on a fresh qubit. Two kinds of bad events
can occur at $v_i$.

First, the correct outcome $k^{*}$ may fail to reach the threshold in the
correct basis $M(k)$. Each measurement in $M(k)$ yields $k^{*}$ with
probability at least $1-e$ (Lemma~\ref{lem:correct-basis}), and
$(1-e)-\tau = \gamma/2$, so by the Chernoff--Hoeffding bound the empirical
frequency of $k^{*}$ falls below $\tau$ with probability at most
$e^{-2n(\gamma/2)^{2}} = e^{-n\gamma^{2}/2}$.

Second, some fixed outcome $j^{*}$ of a wrong basis $M(j)$, $j \neq k$, may
reach the threshold. Each measurement in $M(j)$ yields $j^{*}$ with probability
at most $\delta_{e}$ (Lemma~\ref{lem:wrong-basis}), and
$\tau-\delta_{e} = \gamma/2$, so the empirical frequency of $j^{*}$ reaches
$\tau$ with probability at most $e^{-n\gamma^{2}/2}$.

Applying the union bound over the correct-basis event and the two outcomes of
each of the $\Delta/2-1$ wrong bases,
\[
  \Pr[\text{bad event at } v_i]
  \le \left(1 + 2\!\left(\tfrac{\Delta}{2}-1\right)\right) e^{-n\gamma^{2}/2}
  \le \Delta\, e^{-n\gamma^{2}/2}.
\]
If no bad event occurs, exactly one outcome reaches the threshold, namely
$k^{*}$, and the agent decodes correctly. Choosing
$n = O\!\left(\dfrac{\log D + \log\Delta}{\gamma^{2}}\right)$ gives
$\Delta e^{-n\gamma^{2}/2}=O(1/D)$. Since pebbles are prepared independently,
bad events at distinct nodes are independent, and correctly decoding all $D$
nodes has probability
\[
  \Pr[\text{succ}]
  \ge \left(1-\Delta e^{-n\gamma^{2}/2}\right)^{D}
  = \left(1-O\!\left(\tfrac{1}{D}\right)\right)^{D}
  \longrightarrow e^{-O(1)} > 0,
\]
and choosing $\Delta e^{-n\gamma^{2}/2}=c/D$ for arbitrarily small $c>0$ makes
the success probability arbitrarily close to $1$. \qed
\end{proof}

% =====================================================================
% ---------------------------------------------------------------------
\section{Encoding the Whole Path in a Single Pebble is Not Better}
\label{subsec:single-qubit}
% ---------------------------------------------------------------------
Two distinct variants of the encoding are noted here. Only one of them is
viable.

\paragraph{One pebble per node, higher-dimensional.}
If a $\Delta$-level quantum system (qudit) is available instead of a $2$-level
qubit, then the port number \emph{at that node} can be encoded directly in the
computational basis and recovered with a single measurement, removing the need
to search over $\Delta/2$ bases. This is a genuine improvement on the protocol
of Section~\ref{subsec:gen-protocol}; it still places one pebble per node.

\paragraph{One pebble in total, encoding the whole path.}
One might instead hope to encode the \emph{entire} sequence of port numbers in
the pebble at the source node and place no further pebbles, so that the agent
measures only once at the start. This variant is not viable, for two
independent reasons.

First, it is incompatible with obliviousness. An agent that has
decoded the whole path $e_1,\dots,e_D$ at the source must still know
\emph{which} of these ports to use at the node it currently occupies. That
requires remembering how many steps it has taken, which is exactly the
persistent state an oblivious agent does not have
(Definition~\ref{def:oblivious}); and after leaving the source there is no
pebble to re-read. Thus no oblivious agent can exploit a whole-path encoding,
whatever its dimension.

Second, even setting obliviousness aside and granting the agent the ability to
index into the decoded path, the encoding is exponentially more expensive to
read. We note this quantitatively. It is the number of bases
that drives the measurement cost. There are $\Delta^{D}$ candidate paths,
hence $\Delta^{D}/2$ bases, and the analogue of Lemma~\ref{lemma:angle}
degrades accordingly.

\begin{lemma}
\label{lemma:angle-full}
For $j \neq k$ and $* \in \{+,-\}$,
$\ |\braket{j_*}{k_*}|^2 \le \cos^2\!\left(\dfrac{\pi}{2\Delta^{D}}\right).$
\end{lemma}
\begin{proof}
The proof is the same as that of Lemma~\ref{lemma:angle}, with $\Delta^{D}$ in
place of $\Delta$ throughout: the basis index now satisfies
$k-j \le \Delta^{D}/2-1$ and the offset is $\phi = \pi/\Delta^{D}$.
\end{proof}

%The bound of Lemma~\ref{lemma:angle-full} is exponentially closer to $1$ than
%that of Lemma~\ref{lemma:angle}, since exponentially more states must be
%distinguished. 

Let $\delta_{D} := \cos^2(\pi/2\Delta^{D})$. Repeating the
union-bound argument of Theorem~\ref{thm:main}, now over the
$\Delta^{D}/2 - 1$ wrong bases, the number $n_{1}$ of measurements at the
first node must satisfy $\Delta^{D}\delta_{D}^{\,n_{1}} = O(1)$, i.e.
\[
  n_{1} \;=\; \Omega\!\left(\frac{D\log\Delta}{\log(1/\delta_{D})}\right).
\]
Since $-\ln(1-x) \le 2x$ for $x \in [0,\tfrac12]$ and
$1-\delta_{D} = \sin^{2}(\pi/2\Delta^{D}) \le (\pi/2\Delta^{D})^{2}$, we have
$\log(1/\delta_{D}) \le \pi^{2}/(2\Delta^{2D})$, and therefore
\[
  n_{1} \;=\; \Omega\!\left(\Delta^{2D}\, D\log\Delta\right).
\]
{\sloppy In contrast, the per-node scheme of Theorem~\ref{thm:main} uses
$D \cdot O((\log D + \log\Delta)/\log(1/\delta))$ measurements
in total. As $\log(1/\delta) \ge 1-\delta = \sin^{2}(\pi/2\Delta)$ and
$\sin^{2}(\pi/2\Delta) \ge 1/\Delta^{2}$ (using $\sin x \ge 2x/\pi$ on
$[0,\pi/2]$), this total is
$O\!\left(\Delta^{2} D(\log D + \log\Delta)\right)$. The whole-path scheme
therefore requires a number of measurements exponential in $D$, whereas the
per-node scheme is polynomial in $D$. Hence, even for an agent with memory,
encoding the whole path in a single pebble is strictly worse than encoding the
next step at each node.\par}

% =====================================================================

% =====================================================================
\section{Discussion and Further Work}
\label{sec:discussion}
% =====================================================================
%If a $\Delta$-level quantum system (qudit) is available instead of a $2$-level
%qubit, the port numbers can be encoded directly in the computational basis and
%recovered with a single measurement.
A $c$-coloured pebble is deterministically prepared and therefore deterministically read. Therefore repetition adds 
no new information, and  $\log (c+1)$ bits of information is a hard ceiling. A quantum pebble is also deterministically 
prepared, yet measurement regenerates  random information on every
read, so a device carrying at most one bit per read can nevertheless communicate
all $\log{\Delta}$ bits. No classical pebble with fewer than $\Delta-1$ colours has
this property, no matter how many times it is read.

The quantum advantage established in this paper is not a sample
complexity advantage over all classical approaches. A stochastic emitter that returns i.i.d.\ samples with mean $k/(\Delta-1)$ (for port $k$) has the same features as quantum pebbles.
What distinguishes the quantum pebble from the stochastic emitter is {\it qualitative}:
the quantum pebble is deterministically prepared (a pure state, with
no randomness stored in the device), and unlike classical broadcasts, cannot be cloned. 
Whether no-cloning yields a quantitative separation from stochastic emitters in an adversarial 
 variant of treasure hunt is the object of a future study.

Two natural improvements are possible. First, the current protocol with no state preparation error
always takes $n$
measurements per basis. The average running time can improve by stopping once the
measurements in a basis cease to give a uniform string. One can possibly quantify this gain in the
running time. Second, one can eliminate the
failure condition entirely as follows. Rotate through all bases, maintaining the running
measurement string for each, discard any basis that shows a non-uniform run, and
advance to the next node once all but one basis has been eliminated. The gain needs to be quantified for this modification.

Lemma~\ref{lemma:angle} used a weak lower bound of $\pi/2\Delta$ on the angle
between bases $M(i)$ and $M(j)$. The  angles are in fact a multiple of
$\pi/2\Delta$, so the bound in Theorem~\ref{thm:main} can be refined, likely
reducing the dependence on $\Delta$ in the number of measurements per node. Relatedly, a $\Delta$-level qudit at each node would allow the port to be read
with a single measurement in the computational basis, avoiding the search over
$\Delta/2$ bases. Encoding the entire path in one pebble
instead, as discussed in Section~\ref{subsec:single-qubit}, is incompatible with
obliviousness and in any case exponentially more expensive to read. For the robustness results of Section~\ref{sec:robustness}, a
tighter characterization of $e^{*}$ under structured noise, and experimental
validation, is an interesting question.

% =====================================================================
\section{Conclusion}
\label{sec:conclusion}
% =====================================================================
This paper investigates treasure hunt in anonymous graphs by oblivious agents,
where nodes are unlabelled, edges carry only local port numbers, and agents have
no memory across rounds. Classical uniform pebbles cannot guide such agents, and
randomized classical strategies are ineffective. We replace classical pebbles
with \emph{quantum pebbles} placed along the shortest path, each emitting qubits
that encode the next port; the oblivious agent measures in several bases to
recover the port to follow.

The quantum protocol finds the treasure in $O(D)$ steps using $D$ quantum
pebbles, {\sloppy with $O((\log D + \log\Delta)/\log(1/\delta))$ measurements per
basis, that is, $O(\Delta^{3}(\log D + \log\Delta))$ per node. Its
distinguishing feature is robustness. A single mislabelled classical
pebble traps the agent forever. Furthermore, no randomized classical strategy escapes
exponential decay. The quantum protocol self-corrects in presence of state preparation
errors through additional
sampling and succeeds with probability approaching $1$ for
$e<e^{*}=\sin^2(\pi/2\Delta)/(1+\sin^2(\pi/2\Delta))$. 

The contributions are:
(i) proposal to use quantum pebbles as a guidance mechanism; (ii) an encoding of directional
information into the state of a single qubit; (iii) measurement in different
bases to extract it; and (iv) a proof that the strategy finds the treasure in $D$ steps with probability close to 1. This is in sharp contrast with the classical impossibility results and the
inefficiency of randomized classical walks. (v) The protocol proposed here is resilient to per-emission
state preparation errors, where the classical protocol fails. Under per-node
persistent errors, however, both paradigms, classical and quantum, fail.
(Proposition~\ref{prop:quantum-persistent}). So the robustness is due
to the regenerative nature of the emitter.\par}
%(Section~\ref{subsec:single-qubit}).

\begin{credits}
\subsubsection{\ackname}
We thank the anonymous reviewers for their careful reading and, in particular,
for suggesting the distinction between per-node persistent and
per-emission preparation errors, which led to
Proposition~\ref{prop:quantum-persistent} and to a more precise statement of
the scope of our robustness result.
Work supported by Natural Sciences and Engineering Research Council of Canada,
Discovery Grant RGPIN-2021-03828 and Anusandhan National Research
Foundation~(ANRF) of India, under Core Research Grant CRG/2023/007610.
\end{credits}

\bibliographystyle{splncs04}
\bibliography{mybibliography}

@inproceedings{akbaristoc25,
author = {Akbari, Amirreza and Coiteux-Roy, Xavier and d'Amore, Francesco and Le Gall, Fran\c{c}ois and Lievonen, Henrik and Melnyk, Darya and Modanese, Augusto and Pai, Shreyas and Renou, Marc-Olivier and Rozho\v{n}, V\'{a}clav and Suomela, Jukka},
title = {Online Locality Meets Distributed Quantum Computing},
year = {2025},
isbn = {9798400715105},
publisher = {Association for Computing Machinery},
address = {New York, NY, USA},
url = {https://doi.org/10.1145/3717823.3718211},
doi = {10.1145/3717823.3718211},
booktitle = {Proceedings of the 57th Annual ACM Symposium on Theory of Computing},
pages = {1295–1306},
numpages = {12},
location = {Prague, Czechia},
series = {STOC '25}
}

@inbook{Balliusoda26,
author = {Alkida Balliu and Filippo Casagrande and Francesco d’Amore and Massimo Equi and Barbara Keller and Henrik Lievonen and Dennis Olivetti and Gustav Schmid and Jukka Suomela},
title = {Distributed Quantum Advantage in Locally Checkable Labeling Problems},
booktitle = {Proceedings of the 2026 Annual ACM-SIAM Symposium on Discrete Algorithms (SODA)},
chapter = {},
pages = {1268-1308},
doi = {10.1137/1.9781611978971.49},
URL = {https://epubs.siam.org/doi/abs/10.1137/1.9781611978971.49},
eprint = {https://epubs.siam.org/doi/pdf/10.1137/1.9781611978971.49}
}

@article{flocchini2013computing,
  title={Computing without communicating: Ring exploration by asynchronous oblivious robots},
  author={Flocchini, Paola and Ilcinkas, David and Pelc, Andrzej and Santoro, Nicola},
  journal={Algorithmica},
  volume={65},
  pages={562--583},
  year={2013},
  publisher={Springer}
}

@article{raussendorf2001one,
  title={A one-way quantum computer},
  author={Raussendorf, Robert and Briegel, Hans J},
  journal={Physical review letters},
  volume={86},
  number={22},
  pages={5188},
  year={2001},
  publisher={APS}
}

@article{briegel2009measurement,
  title={Measurement-based quantum computation},
  author={Briegel, Hans J and Browne, David E and D{\"u}r, Wolfgang and Raussendorf, Robert and Van den Nest, Maarten},
  journal={Nature Physics},
  volume={5},
  number={1},
  pages={19--26},
  year={2009},
  publisher={Nature Publishing Group UK London}
}

@article{nielsen2003quantum,
  title={Quantum computation by measurement and quantum memory},
  author={Nielsen, Michael A},
  journal={Physics Letters A},
  volume={308},
  number={2-3},
  pages={96--100},
  year={2003},
  publisher={Elsevier}
}

@article{nielsen1997programmable,
  title={Programmable quantum gate arrays},
  author={Nielsen, Michael A and Chuang, Isaac L},
  journal={Physical Review Letters},
  volume={79},
  number={2},
  pages={321},
  year={1997},
  publisher={APS}
}

@book{nielsen2010quantum,
  title={Quantum computation and quantum information},
  author={Nielsen, Michael A and Chuang, Isaac L},
  year={2010},
  publisher={Cambridge university press}
}

@article{gorain2022pebble,
  title={Pebble guided optimal treasure hunt in anonymous graphs},
  author={Gorain, Barun and Mondal, Kaushik and Nayak, Himadri and Pandit, Supantha},
  journal={Theoretical Computer Science},
  volume={922},
  pages={61--80},
  year={2022},
  publisher={Elsevier}
}

@inproceedings{xin2007faster,
  title={Faster treasure hunt and better strongly universal exploration sequences},
  author={Xin, Qin},
  booktitle={International Symposium on Algorithms and Computation},
  pages={549--560},
  year={2007},
  organization={Springer}
}

@inproceedings{langetepe2010optimality,
  title={On the optimality of spiral search},
  author={Langetepe, Elmar},
  booktitle={Proceedings of the twenty-first annual ACM-SIAM symposium on Discrete Algorithms},
  pages={1--12},
  year={2010},
  organization={SIAM}
}

@inproceedings{bose2013revisiting,
  title={Revisiting the problem of searching on a line},
  author={Bose, Prosenjit and De Carufel, Jean-Lou and Durocher, Stephane},
  booktitle={Algorithms--ESA 2013: 21st Annual European Symposium, Sophia Antipolis, France, September 2-4, 2013. Proceedings 21},
  pages={205--216},
  year={2013},
  organization={Springer}
}

@article{kao1996searching,
  title={Searching in an unknown environment: An optimal randomized algorithm for the cow-path problem},
  author={Kao, Ming-Yang and Reif, John H and Tate, Stephen R},
  journal={Information and computation},
  volume={131},
  number={1},
  pages={63--79},
  year={1996},
  publisher={Elsevier}
}

@article{pattanayak2024graph,
  title={Graph exploration by a deterministic memoryless automaton with pebbles},
  author={Pattanayak, Debasish and Pelc, Andrzej},
  journal={Discrete Applied Mathematics},
  volume={356},
  pages={149--160},
  year={2024},
  publisher={Elsevier}
}

@book{alpern2006theory,
  title={The theory of search games and rendezvous},
  author={Alpern, Steve and Gal, Shmuel},
  volume={55},
  year={2006},
  publisher={Springer Science \& Business Media}
}

@article{bouchard2023almost,
  title={Almost-Optimal Deterministic Treasure Hunt in Unweighted Graphs},
  author={Bouchard, S{\'e}bastien and Dieudonn{\'e}, Yoann and Labourel, Arnaud and Pelc, Andrzej},
  journal={ACM Transactions on Algorithms},
  volume={19},
  number={3},
  pages={1--32},
  year={2023},
  publisher={ACM New York, NY}
}

@article{bouchard2020deterministic,
  title={Deterministic treasure hunt in the plane with angular hints},
  author={Bouchard, S{\'e}bastien and Dieudonn{\'e}, Yoann and Pelc, Andrzej and Petit, Franck},
  journal={Algorithmica},
  volume={82},
  pages={3250--3281},
  year={2020},
  publisher={Springer}
}

@article{miller2015tradeoffs,
  title={Tradeoffs between cost and information for rendezvous and treasure hunt},
  author={Miller, Avery and Pelc, Andrzej},
  journal={Journal of Parallel and Distributed Computing},
  volume={83},
  pages={159--167},
  year={2015},
  publisher={Elsevier}
}

@article{pelc2019cost,
  title={Cost vs. information tradeoffs for treasure hunt in the plane},
  author={Pelc, Andrzej and Yadav, Ram Narayan},
  journal={arXiv preprint arXiv:1902.06090},
  year={2019}
}

@article{pelc2018reaching,
  title={Reaching a target in the plane with no information},
  author={Pelc, Andrzej},
  journal={Information Processing Letters},
  volume={140},
  pages={13--17},
  year={2018},
  publisher={Elsevier}
}

@inproceedings{komm2015treasure,
  title={Treasure hunt with advice},
  author={Komm, Dennis and Kr{\'a}lovi{\v{c}}, Rastislav and Kr{\'a}lovi{\v{c}}, Richard and Smula, Jasmin},
  booktitle={International Colloquium on Structural Information and Communication Complexity},
  pages={328--341},
  year={2015},
  organization={Springer}
}

@inproceedings{BampasVolatilePheromone,
  author       = {Evangelos Bampas and
                  Joffroy Beauquier and
                  Janna Burman and
                  William Guy{-}Ob{\'{e}}},
  editor       = {Rotem Oshman},
  title        = {Treasure Hunt with Volatile Pheromones},
  booktitle    = {37th International Symposium on Distributed Computing, {DISC} 2023,
                  October 10-12, 2023, L'Aquila, Italy},
  series       = {LIPIcs},
  volume       = {281},
  pages        = {8:1--8:21},
  year         = {2023},
  doi          = {10.4230/LIPICS.DISC.2023.8}
}

@article{pelc2021advice,
  title={Advice complexity of treasure hunt in geometric terrains},
  author={Pelc, Andrzej and Yadav, Ram Narayan},
  journal={Information and Computation},
  volume={281},
  pages={104705},
  year={2021},
  publisher={Elsevier}
}

@inproceedings{BhattacharyaPebble,
  author       = {Adri Bhattacharya and
                  Barun Gorain and
                  Partha Sarathi Mandal},
  title        = {Pebble Guided Treasure Hunt in Plane},
  booktitle    = {Networked Systems - 11th International Conference, {NETYS} 2023, Benguerir,
                  Morocco},
  series       = {Lecture Notes in Computer Science},
  volume       = {14067},
  pages        = {141--156},
  publisher    = {Springer},
  year         = {2023}
}

\newpage
\end{document}